# Tungsten based metasurface absorber for visible regime

Ahsan Sarwar Rana
Department of Electrical Engineering, ITU
6th Floor, Arfa Software Technology Park,
Ferozepur Road, Lahore
ahsan.sarwar@itu.edu.pk

***Abstract*—utilizing solar energy requires perfect absorption by photovoltaic cells for efficient conversion of solar energy into useful electrical energy. Metasurfaces can, not only be used to absorb solar energy efficiently but, they do so with thickness of structure in nano-scale -effectively reducing the size of the device. To achieve aforementioned feature metasurface absorbers are developed which give better absorbance on narrow/wide frequency spectrum. These absorbers are fashioned with variable structures employing variable materials but still, perfect absorbance is not attained with a simple two-dimensional (not varying in z-direction) structure. This paper presents a simulated realization of a broadband metasurface absorber based on tungsten nano-structure. The design is two-dimensional, polarization insensitive, broadband and is predicted to give better response under high temperatures ascribed to high melting point of tungsten i.e. 3422°C. Cross alignment is found optimum for tungsten as it is impedance matched with the free space for visible spectrum. This cross arrangement is further tweaked by changing span, width and height of the cross resulting in 7 different optimized solutions giving an average absorbance greater than 98%. One, amongst these solutions, gave a maximum average absorbance of 99.3%.**

## I. INTRODUCTION

Modern society relies on power to function which comes from different sources of energy. They can be grouped into two categories as 'renewable' and 'non-renewable'. In the former category sun is one of the major sources of energy, the solar energy.

The solar energy can be captured by various mechanisms but one with the most potential is photovoltaic (PV) cells. Basic function of a PV cell is the generation of charge carriers which are then collected in external circuit generating electric current. For a PV cell to have high efficiency, there is a need of high absorbance of solar spectra i.e. absorbance of photons varying in energies therefore, different wavelengths. This calls for a highly efficient absorber.

The advent of sophisticated deposition techniques allowed construction of subwavelength structures (having lower dimensions than operating wavelength). These metamaterials (MMs) show variable properties than regular materials as they give structures variable refractive index. MMs give the freedom to alter permittivity of a material as a whole, which in turns gives varying refractive index thus achieving a mutable response from the device.

The first mention of a "perfect absorber" came in 2008 [1], which showed a perfect light absorber (PLA) in microwave regime at 11.5 GHz. Same group proposed a perfect absorber in visible region [2]. Since then, many broadband absorbers are sought with different structures employing the phenomenon of surface plasmon resonance or impedance matching [3] for maximizing absorbance over a specific wavelength.

Broadband PLAs are also established with a topology of metal-dielectric-metal layers. Boradband Structures previously demonstrated had different variations of simpler structures such as nano-pillars made up of different variations of cylinders [4], disks (cylinders) [5, 6], arrangement of crosses and cylinders [7], nano-pyramids (cones) [8-10], rods (in shape of square) [11], spheres , tetrahedral structure [12], and even random deposition of particles [13], therefore, a simple structure of cross is used as a design in this research.

## II. TUNGSTEN BASED ABSORBER

In the proposed tungsten based absorber for visible regime, the structure contains a ground plane made up of a metal underneath a dielectric layer which in turn is below a resonating structure made up of a same metal as ground plane (metal-dielectric-metal) as shown in Fig.1. Dielectric layer is made up of silicon dioxide ($SiO_2$) as $SiO_2$ has fairly low relatively permittivity at optical frequency range and this, more or less remains constant. $SiO_2$ also provides fairly high melting point which is also a desired property for the dielectric layer.

This paper is on introduction of tungsten (W) which has higher melting point than any other metal i.e. Gold (Au), Silver (Ag), Chromium (Cr), Copper (Cu) and even Titanium Nitride (TiN) which is a refractory material. Previous structures involving W are made up in the shape of cones as discussed in [9, 10], but the designs presented in this paper are 2D as they are not varying in z-direction (height). The design presented in [6] is also 2D but it is not designed for visible regime and it gives less absorbance. Since tungsten does not support surface plasmons in optical range, high absorbance is achieved by impedance matching to the free space.

A design of square ring structure is taken here as an initial point. After the optimization of square ring structure, it is observed that the structure gives negligible reflectance if the top layer is made in the shape of cross. The cross structure is then analyzed further in detail.

## III. Results and Discussions

### A. *Four rods (square structure) simulations*

Simulations of the structures are performed in a step by step fashion in Lumerical FDTD solutions. Material properties in the software are explored and curve fitting is performed for tungsten (Palik) and SiO2 (Palik) according to experimental values in [14]. Numbers of maximum coefficients are varied against permittivity and 6 coefficients for SiO2 and 15 coefficients for tungsten are taken, as they best fitted the curve for optical domain. A structure of metal-dielectric (spacer)-metal (ground plane-GP) of a square ring is formed initially as a starting point as shown in Fig.1. The dimensions of the structure In Fig.1 are listed in Table.1.

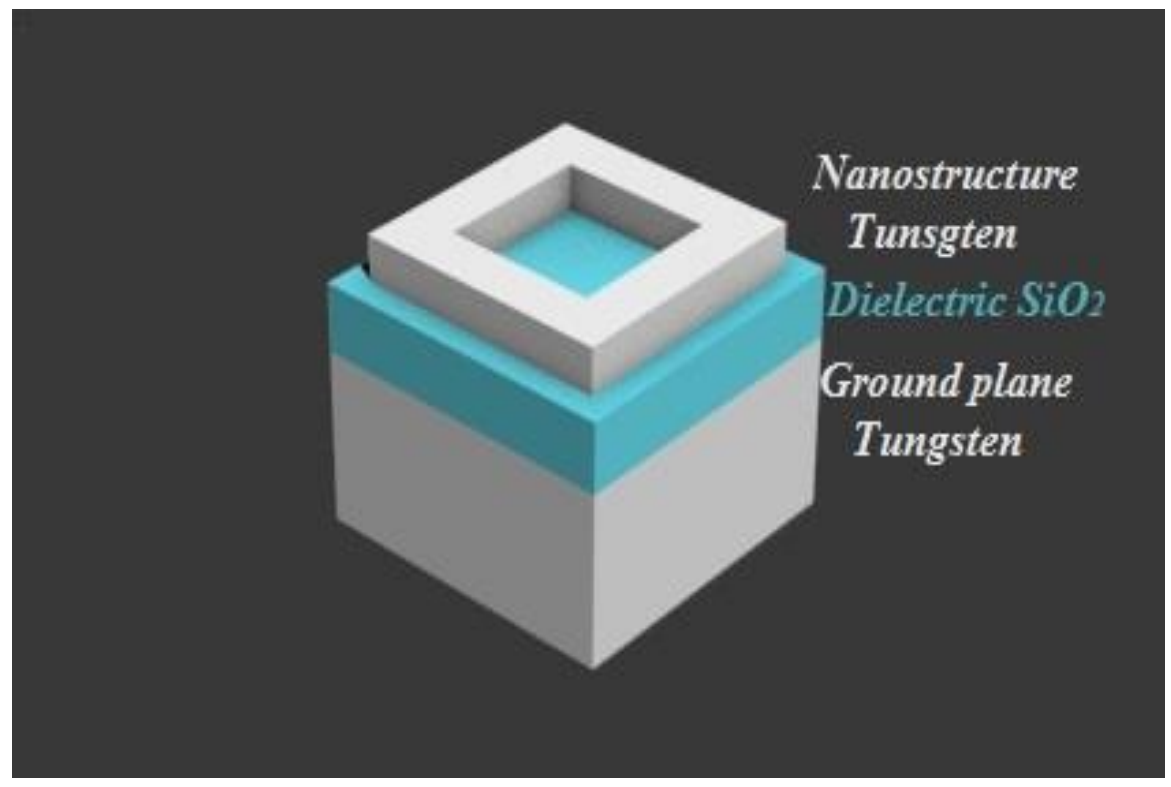

Figure 1: Square ring structure.

Table 1: Dimensions of square ring structure

| Structure | X span (nm) | Y span (nm) | Z span (nm) |
| --- | --- | --- | --- |
| Top rod | 250 | 50 | 40 |
| Right rod | 50 | 250 | 40 |
| Bottom Rod | 250 | 50 | 40 |
| Left Rod | 50 | 250 | 40 |
| Spacer | 300 | 300 | 60 |
| GP | 300 | 300 | 150 |

When the rod is parallel to horizontal, its y span is considered as its width, and when it is perpendicular its x span is width. If the rods are to make a square structure they must be placed at 100 nm if center of the structure is at origin (0, 0). That is to say that the right rod will be at x = 100 nm from origin, left rod at x = -100 nm, top rod at y = 100 nm and bottom rod at y = -100 nm.

The mesh refinement is set to conformal variant 1 as it included metal boundaries. The results shown in this paper are described with three types mesh step settings (Tab.2) to decrease overall time for simulation.

Table 2: Mesh step settings

| Mesh Step Setting | dx | dy | dz |
| --- | --- | --- | --- |
| MSS 1 | 1 | 1 | 5 |
| MSS 2 | 5 | 5 | 5 |
| MSS 3 | 10 | 10 | 10 |

Simulation of this square ring structure resulted in two different absorbance curves for two maximum mesh step settings (MSS 1 and MSS 2) as shown in Fig.2.

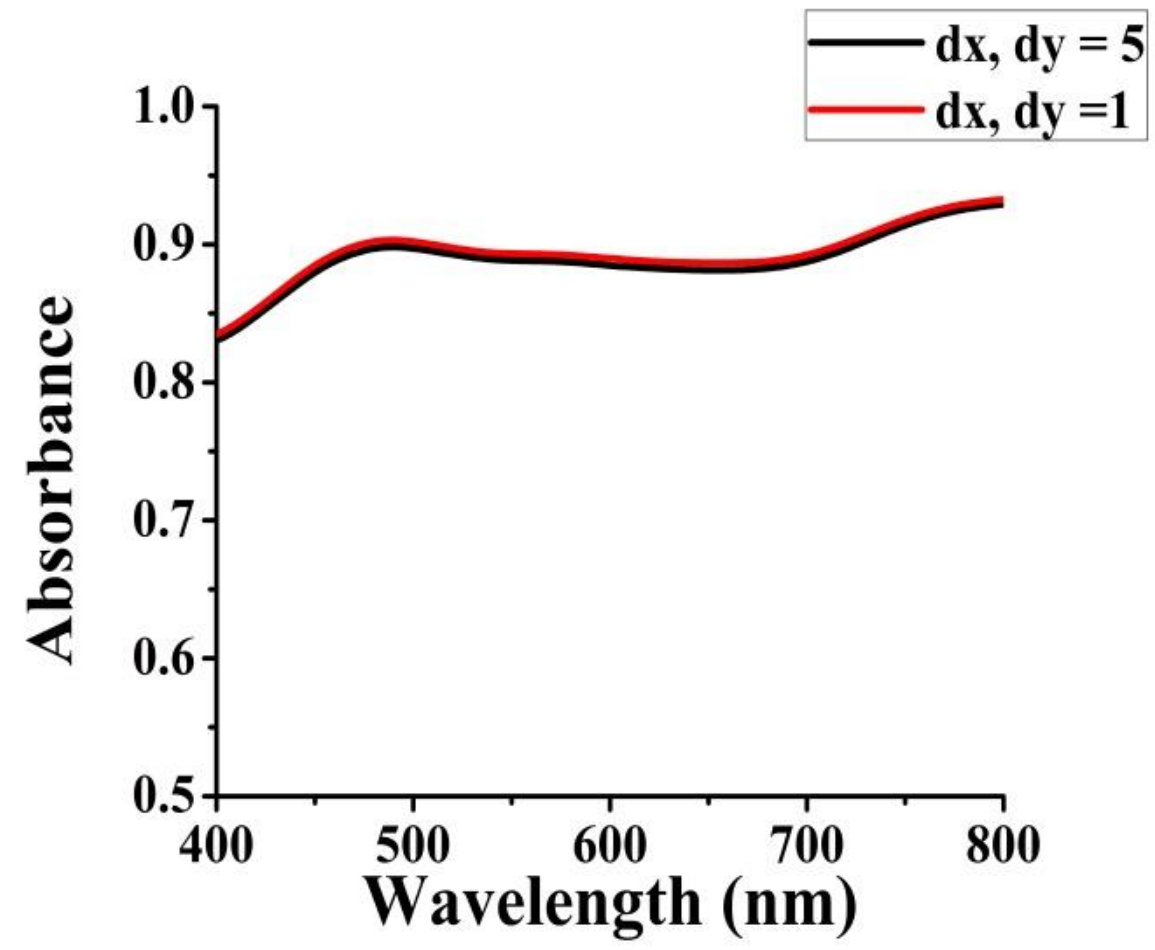

Figure 2: Different mesh step settings for tungsten rods.

The results gives two important conclusions i) the structure is not impedance matched with the free space and ii) the maximum mesh step settings plays a little role in absorbance curve. The results are calculated from equation 1 as simulations give reflection and transmission.

$$A = 1 - |T| - |\Gamma| \quad \text{(i)}$$

Where,
A = Absorbance,
$T$ = Transmission, and
$\Gamma$ = Reflection.

The equation 1 can also be manipulated in terms of s-parameters, given by equation 2.

$$A = 1 - |S_{11}|^2 - |S_{12}|^2 \quad \text{(ii)}$$

### B. *Optimiziation of rods for tungsten*

As the structure of rods os not impedance matched for tungsten, variations are made keeping the structure polarization insensitive. Upon evaluation, it is observed that if the rods are moved towards their opposite rod (top rod towards bottom and vice versa, and similarly, right rod towards left rod and vice versa), the structure still remains polarization insensitive. By changing the top structure, the device impedance varies as a whole and reflection is observed with displacement over the entire optical range in Fig.3.

The rods are moved to the origin (0, 0) and not taken to -100 nm (in case of left and bottom rods to 100), because by symmetry that would result in the same plot (inverted along the horizontal). The results show that structure gives minimum reflection as a cross configuration. Therefore the optimum absorber for tungsten is made into a cross in Fig.4.

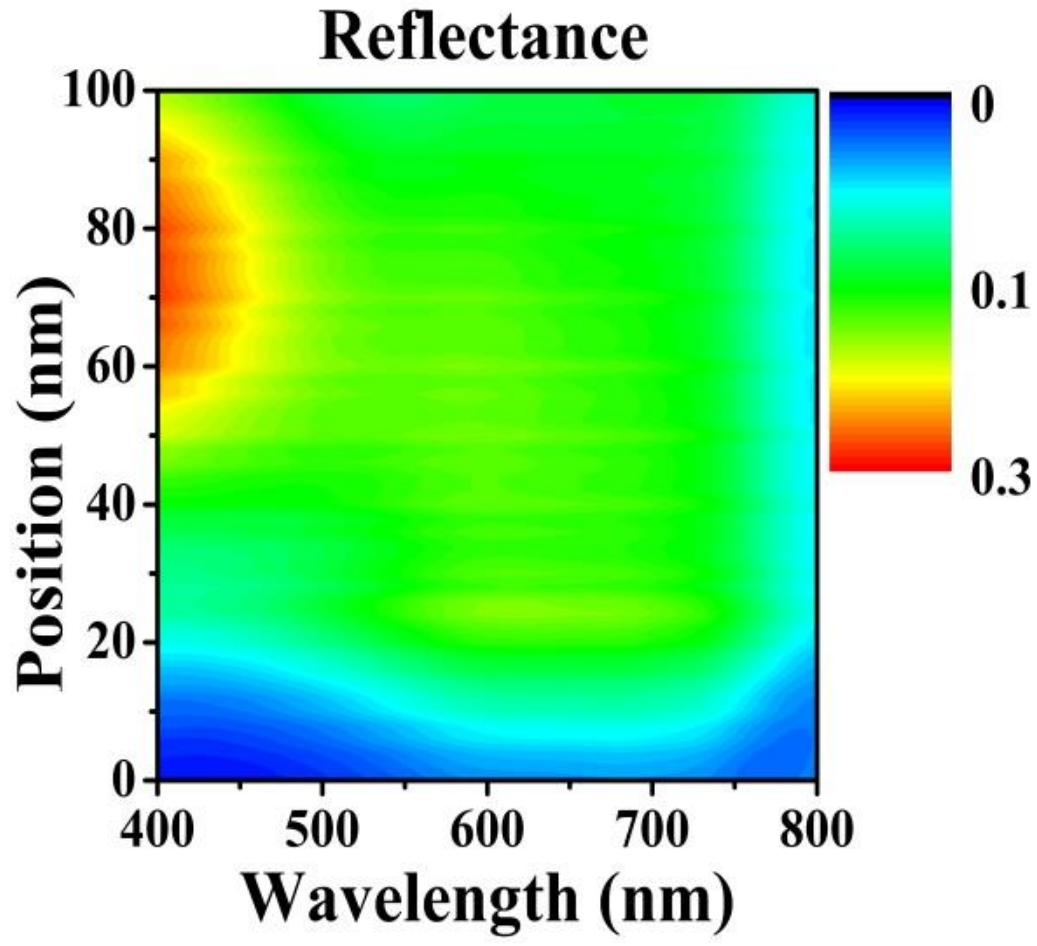

Figure 3: Position variance to get minimum reflection.

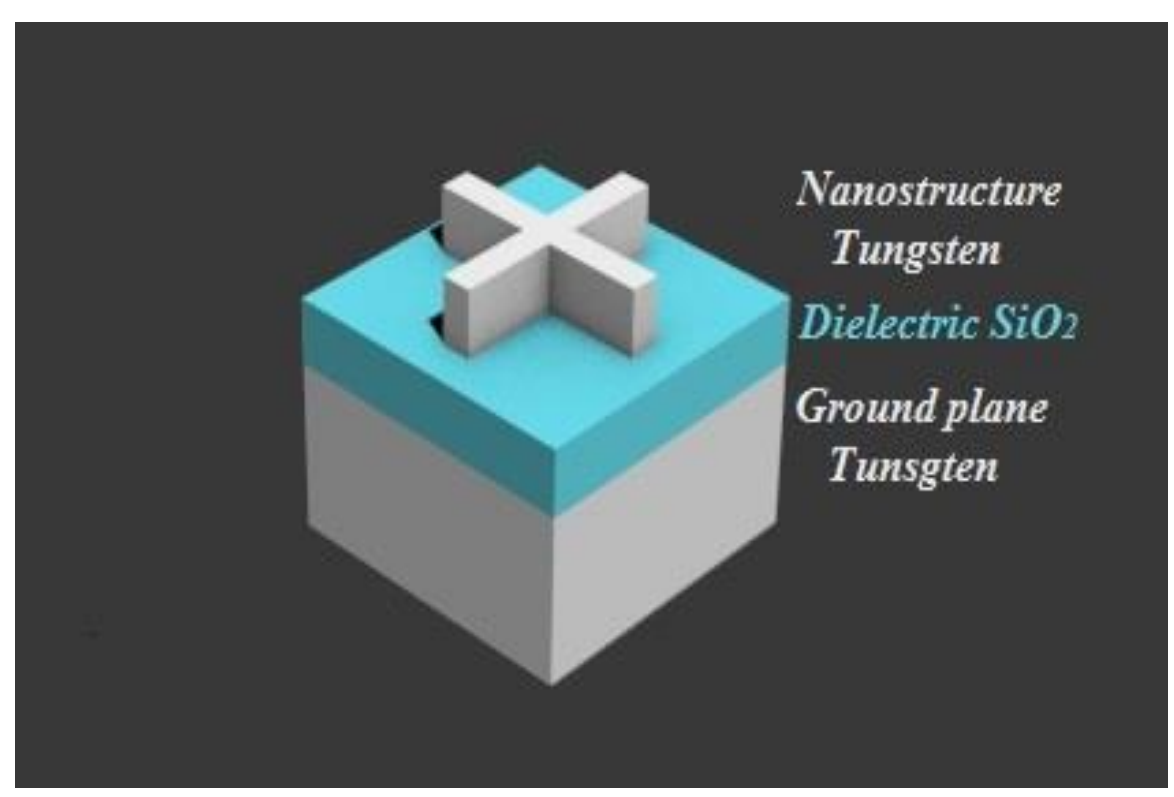

Figure 4: Cross design for tungsten.

The cross configuration can have different variations. It can be varied in width, height and span to change overall impedance of device. Variations in cross structure are made by changing aforementioned parameters and results of absorbance are attained in Fig.5. Using these results average absorbance of the variations is calculated by summation of all points divided by number of points as shown in Table.3. MSS 2 is used for attaining results in Fig.5 and only the best results are further selected with 98% absorbance as a threshold.

Table 3: Different simulation results

| # | Structure dimensions | | | Absorbance (%) |
|---|---|---|---|---|
| | Width (nm) | Height (nm) | Span (nm) | |
| Var1 | 30 | 60 | 200 | 99.2615 |
| Var2 | 30 | 60 | 225 | 99.3141 |
| Var3 | 30 | 80 | 200 | 98.8946 |
| Var4 | 30 | 80 | 225 | 98.9047 |
| Var5 | 40 | 60 | 175 | 98.8504 |
| Var6 | 40 | 60 | 200 | 99.0928 |
| Var7 | 40 | 60 | 225 | 98.5422 |

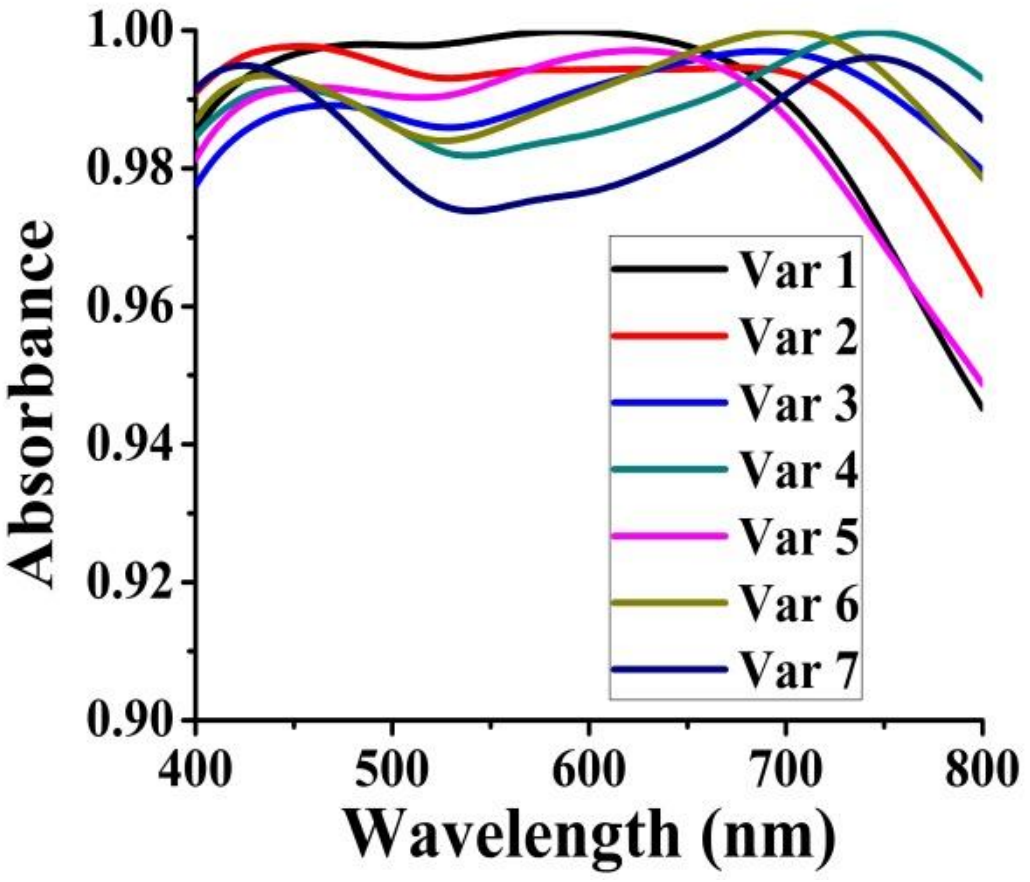

Figure 5: Different variations of cross design.

Different mesh step setting are applied to the design giving best absorbance i.e. Variation 2 and absorbance curves are attained to see if MSS 2 is giving comparable results as shown in Fig.6.

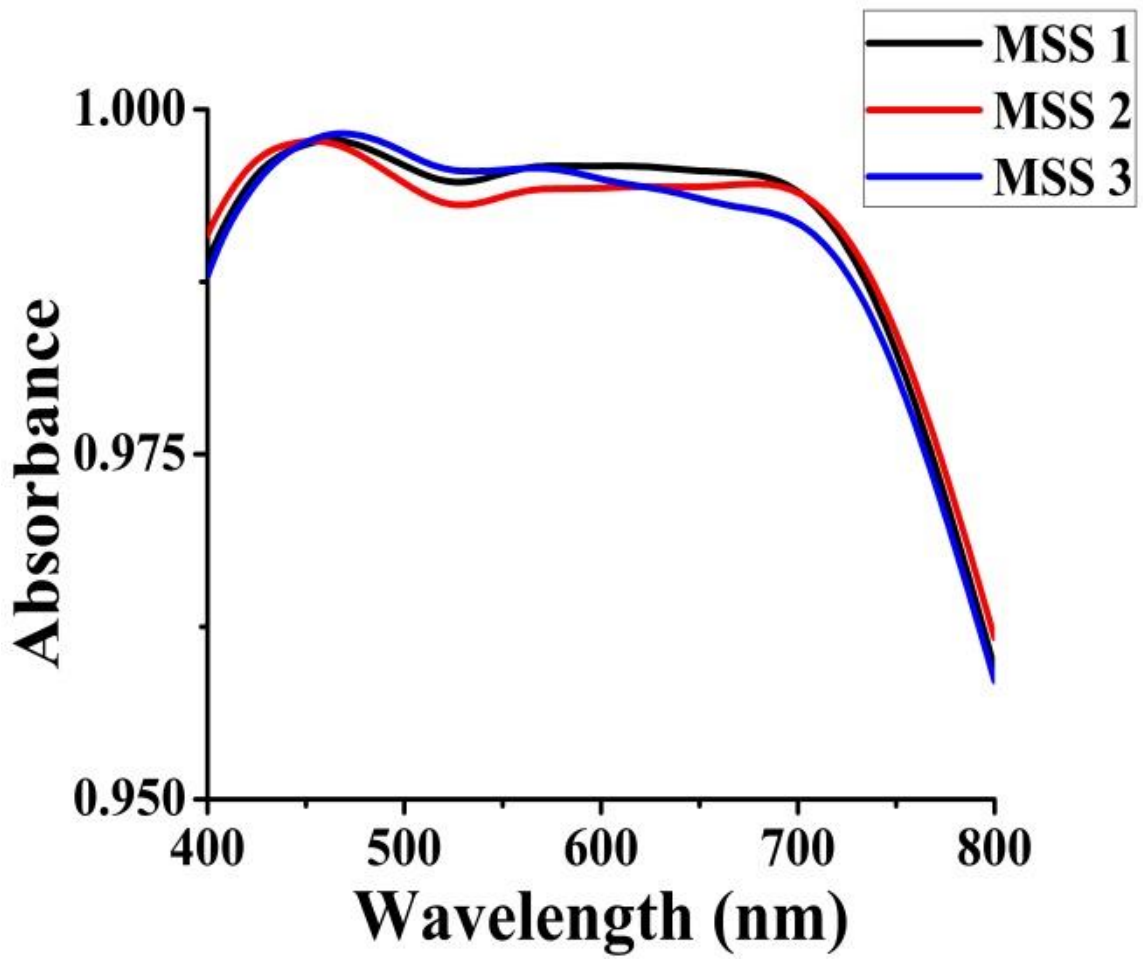

Figure 6: MSS configurations for cross Variation 2.

The results in Fig.6 show an average absorbance of 99.3455%, 99.3141% and 99.2943% for MSS 1, 2 and 3 respectively. This shows that the MSS 1, 2 and 3 can be used interchangeably as they do not produce significant error but using greater step settings reduces the simulation time. Therefore, for the rest of the results paper MSS 2 is used.

Electric-field monitored from just above the cross structure in x-y plane and from a cut-plane of x-z in the center of the structure, Electric-field profile variation of this structure is observed, shown in Fig.7 and Fig.8 respectively. These field profiles suggest a strong field-localization on cross for λ = 800 nm which decreases as the wavelength decreases. Same pattern of field-localization is observed on the rest of the structure.

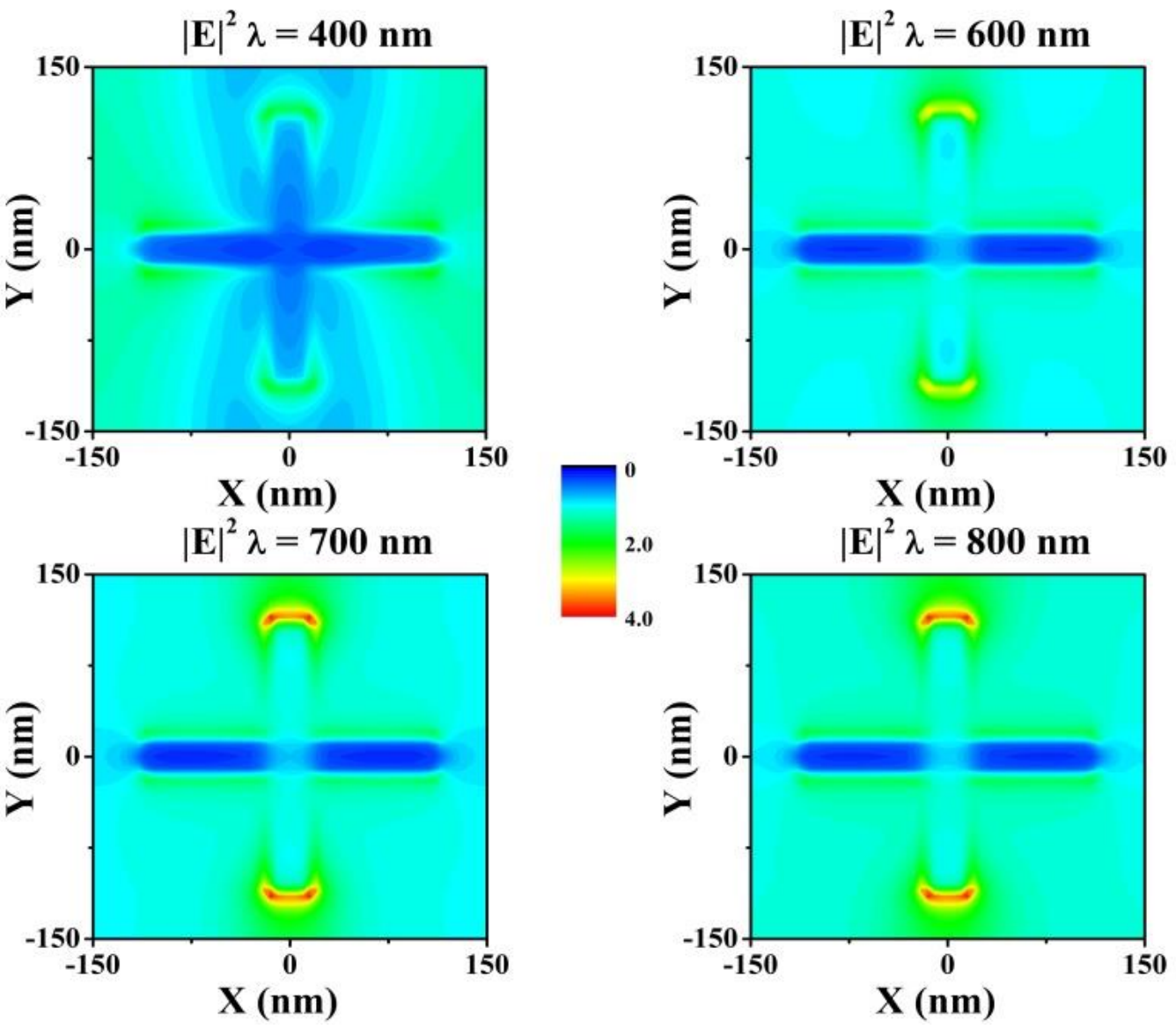

Figure 7: Electric-field profile in x-y plane.

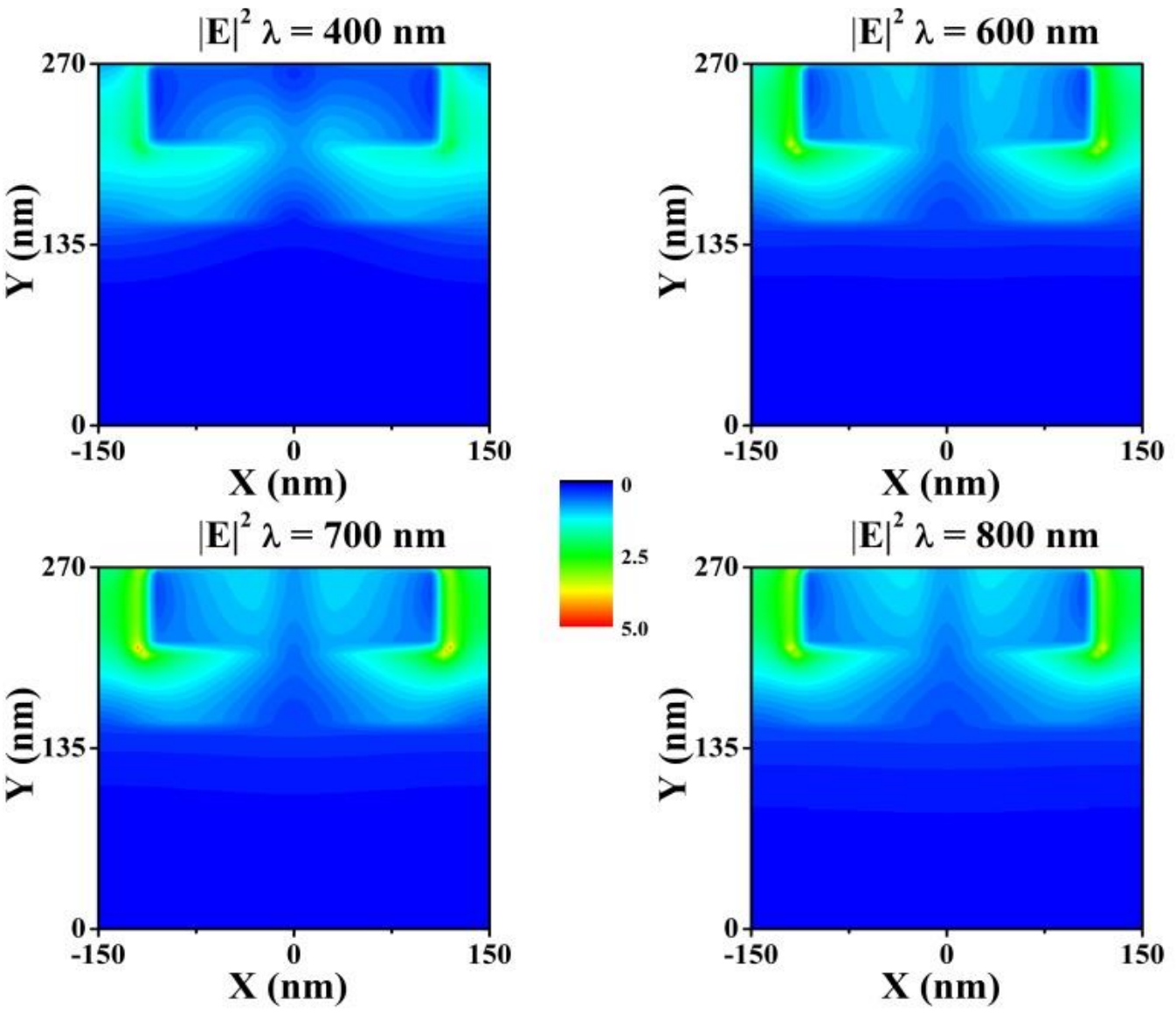

Figure 8: Electric-field profile in x-z plane.

Contributions of front layer and back layer to the absorbance are found out by simulating them separately. If the absorbance from one layer and total absorbance is known, then absorbance from unknown layer is calculated by subtracting the two. The results in Fig.9 show that magnitude of contribution from both layers is quite high. Still, back layer is contributing more towards overall absorbance till λ = 575 nm but after that front layer's contribution is more significant.

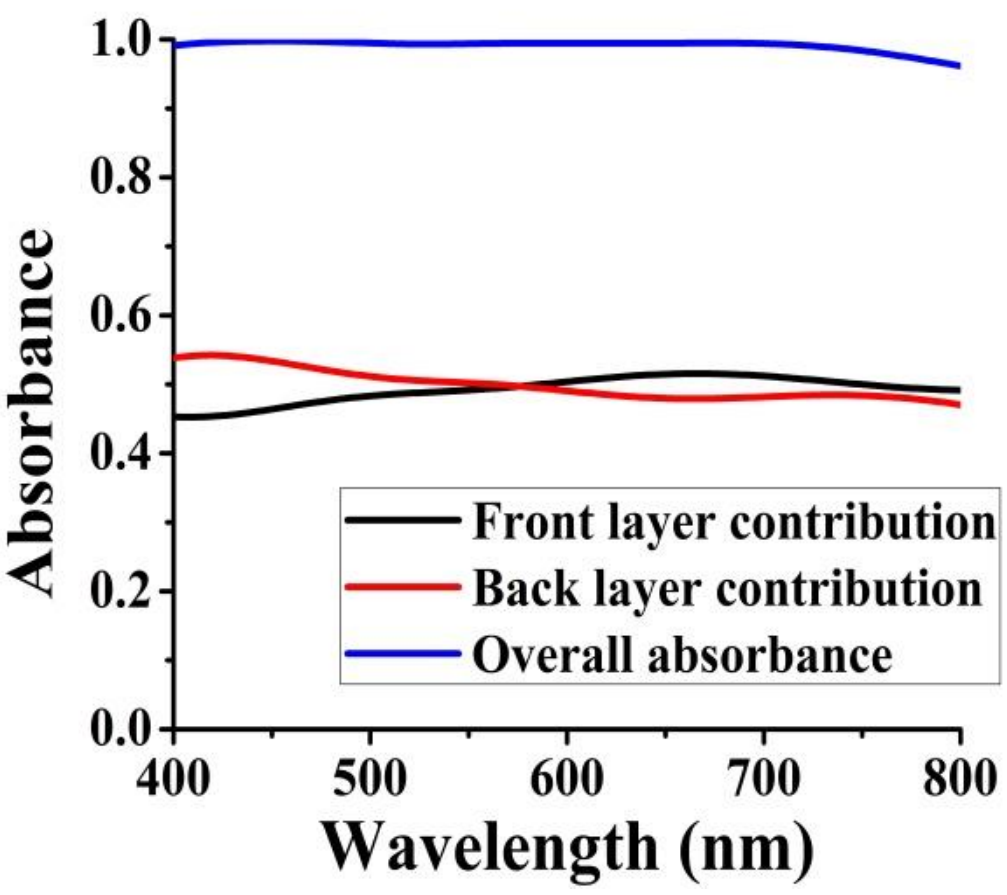

Figure 9: Layer contribution for absorbance.

Such high level of absorbance is achieved because the whole device is impedance matched with the free space. A theoretical proof of this observation is can be attained numerically if the s-parameters of the device are known. The s-parameters for the simulation can be found by using an analysis group in Lumerical FDTD software. After calculating the s-parameters the impedance, refractive index, relative permittivity and relative permeability of the device can be calculated by using the equations [15-18] mentioned below.

$$z = \pm \sqrt{\frac{\left(1+S_{11}^2\right)^2 - S_{21}^2}{\left(1-S_{11}^2\right)^2 - S_{21}^2}} \qquad \text{(iii)}$$

Where,
z = impedance.

$$e^{ink_o d} = X \pm \sqrt{1 - X^2} \qquad \text{(iv)}$$

Where,
$X = \frac{1}{2S_{21}(1-{S_{11}}^2+{S_{21}}^2)}$,
$k_o = \frac{2\pi}{\lambda}$ (The wave number of the free space), and
$d$ = The thickness of the absorber (its height).

$$n = \frac{j(\ln((e^{ink_o d})') + j(e^{ink_o d})'')}{k_o d} \qquad \text{(v)}$$

Where,
n = refractive index,
j = imaginary number,
x' = real part of x, and
x" = imaginary part of x.

The values of impedance and refractive index are chosen such that real value of impedance is greater than 0 and imaginary value of refractive index is greater than 0. The values of effective permittivity and permeability can then be found out by following formulas [15, 17].

$$\varepsilon_r = \frac{n}{z} \quad \text{(vi)}$$

$$\mu_r = nz \quad \text{(vii)}$$

Where,
$\varepsilon_r$ = effective permittivity, and
$\mu_r$ = effective permeability.

The s-parameters from the tungsten cross structure are plotted in Fig.10 and Fig.11, real and imaginary respectively. From there the figures of merits such as z, n, $\varepsilon_r$ and $\mu_r$ are calculated using equations (iii) to (vii) which are then plotted in Fig12 and Fig.13, real and imaginary respectively.

The simulation and manual calculations yield the same result. The s-parameters can also help to find the overall reflection and transmission and therefore, absorbance by using equation (ii). From this analysis it can be safely assumed that reflection and transmission are a result of square of absolute of $S_{11}$ and $S_{21}$ parameters, respectively.

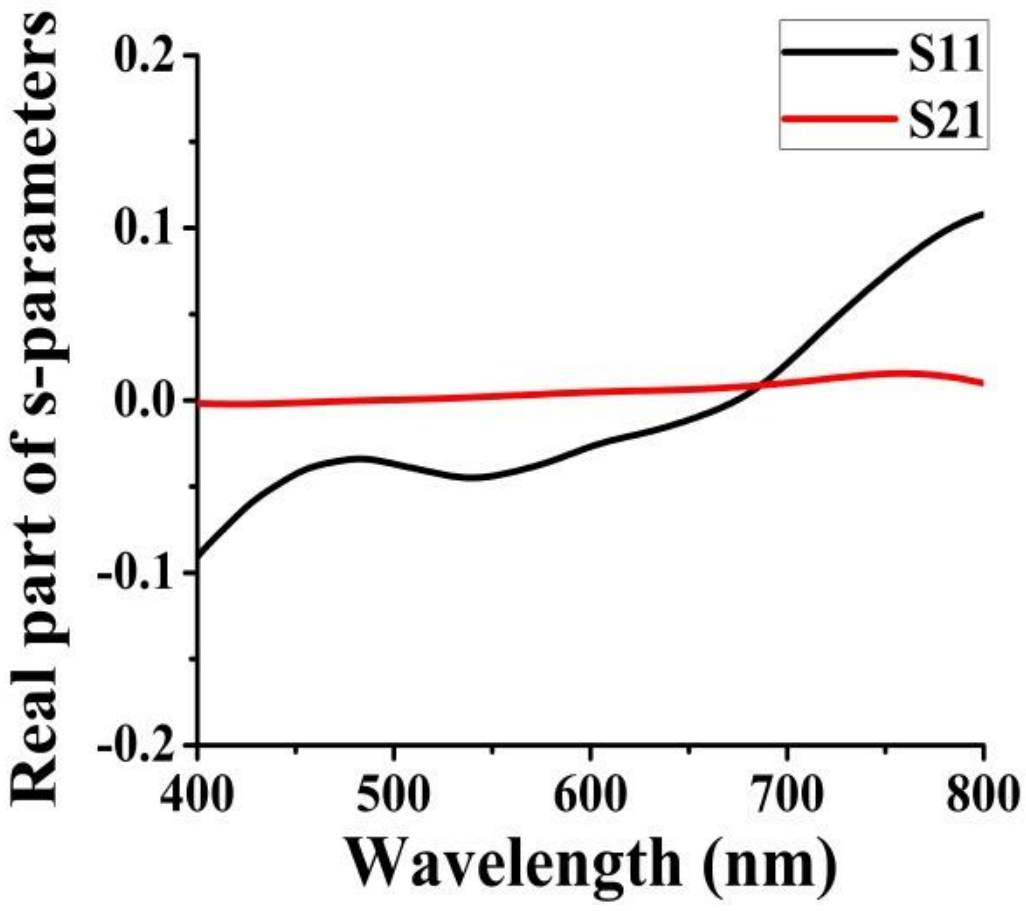

Figure 10: Real part of s-parameters.

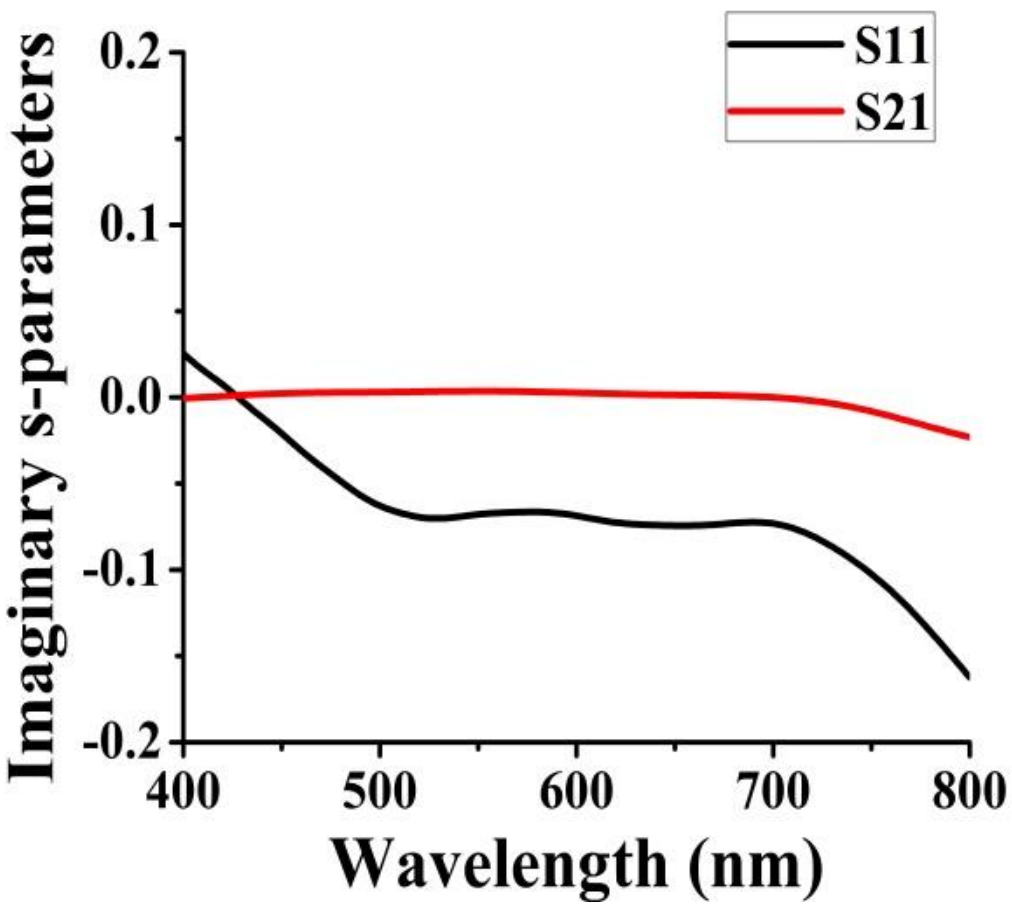

Figure 11: Imaginary part of s-parameters

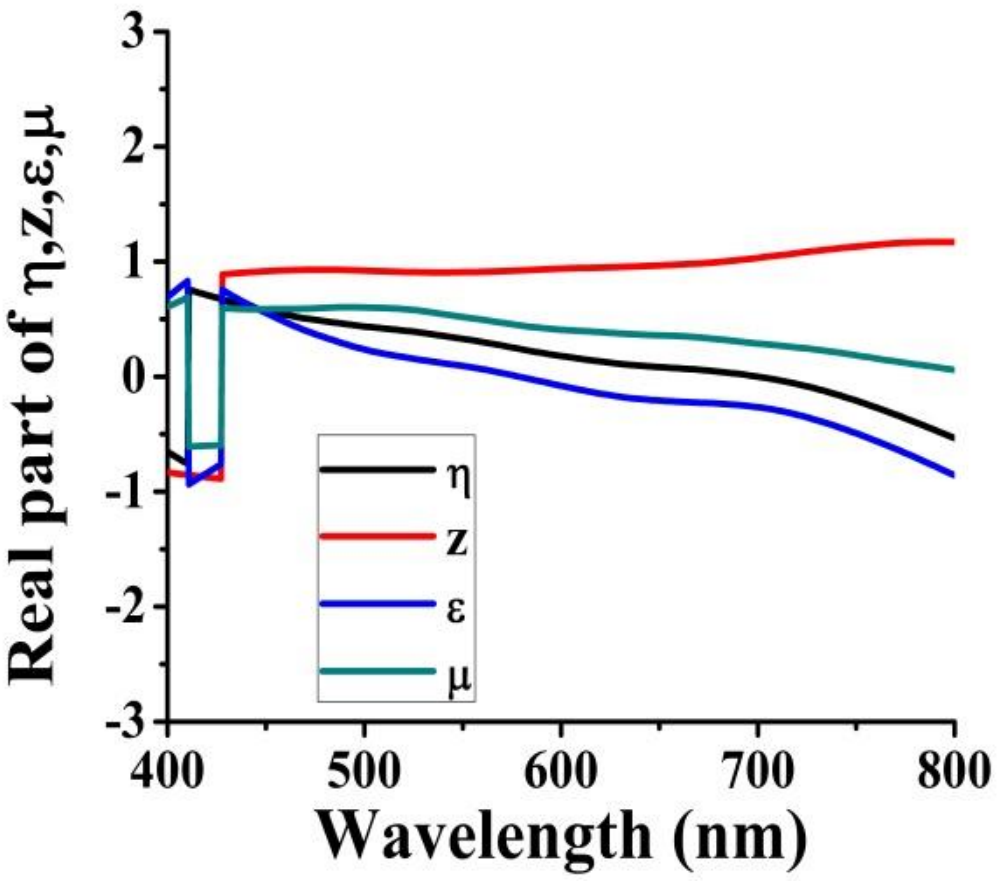

Figure 12: Real part of figures of merits.

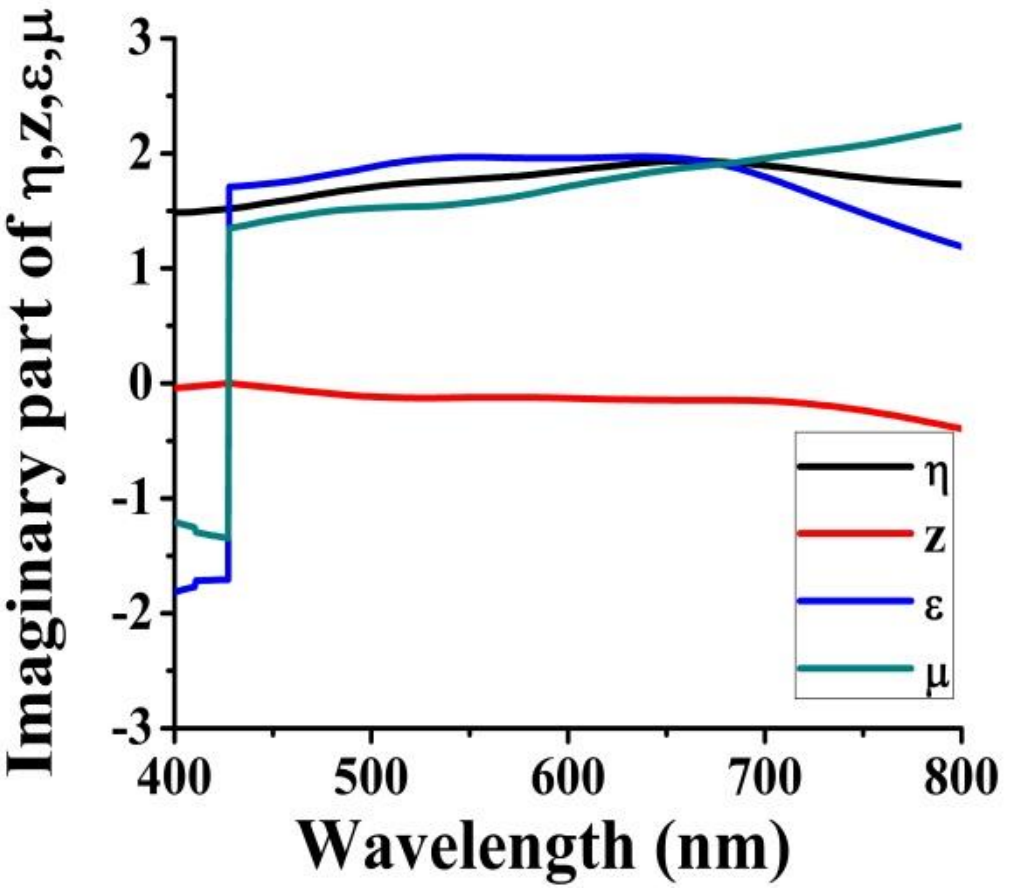

Figure 13: Imaginary part of figures of merits.

As a good absorber, a device should be able to handle variations in angle of incident beam. Therefore, the cross shaped absorber is simulated for s-polarized and p-polarized light to find out the variation in absorbance. The results of angle of incidence θ (theta) with respect to wavelengths in visible regime are shown in Fig.14 and Fig.15, s-polarized and p-polarized respectively. The results are calculated using MSS 3 as it reduced the simulation time exponentially.

The results in Fig.14 and Fig.15 show that the structure is highly optimized for sources, s and p polarized, with a greater angle of incidence. The structure gives almost a unity absorbance for angles as high as 70° ($\theta<70°$). For angles greater than 70° ($\theta>70°$) the design loses its perfect absorbance in both cases of s and p polarized source.

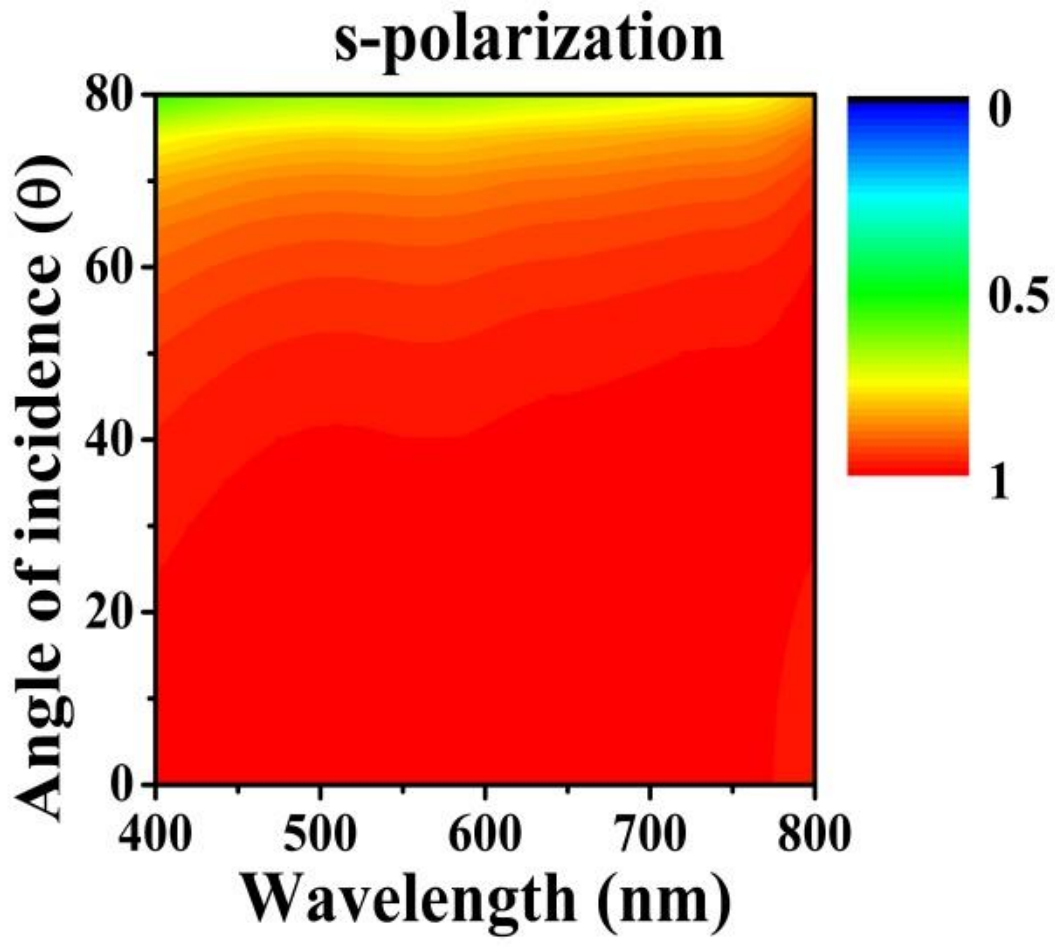

Figure 14: Angle of incidence versus wavelength for s-polarized source.

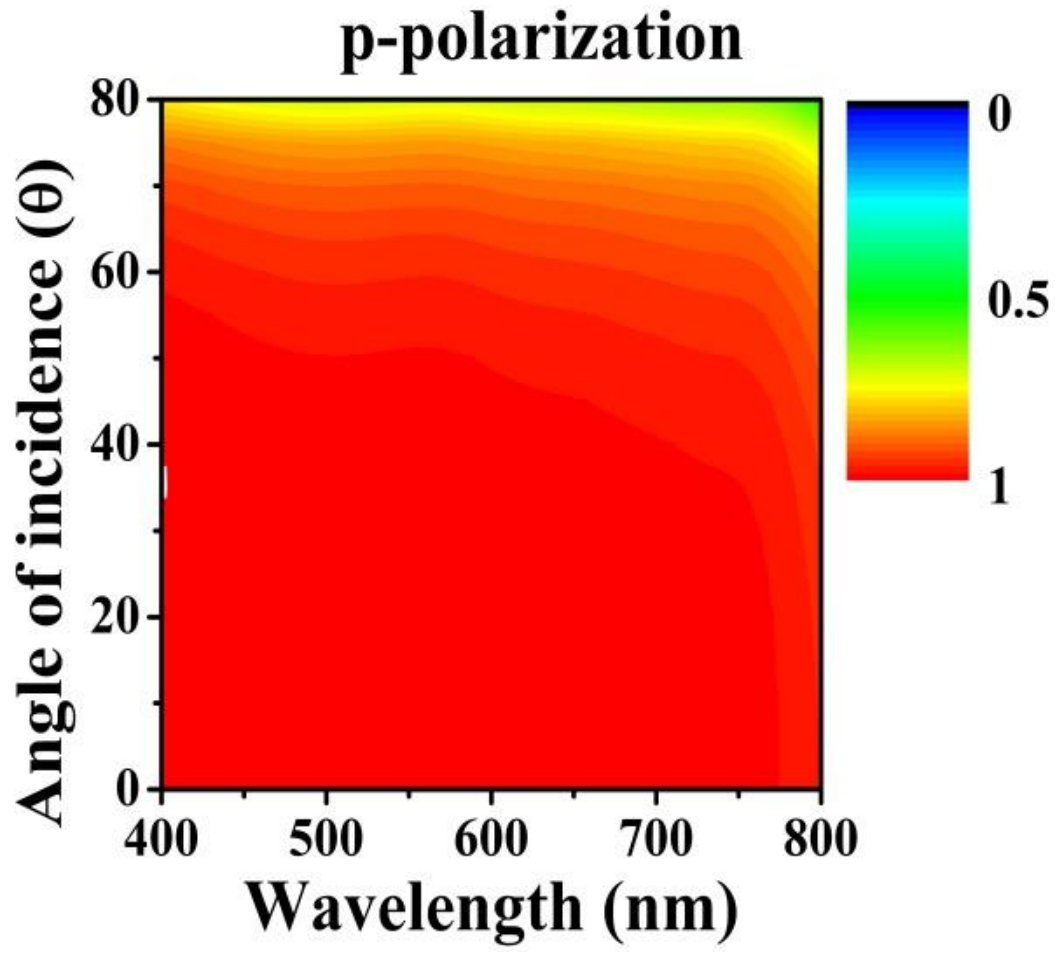

Figure 15: Angle of incidence versus wavelength for p-polarized source.

This structure is also simulated for different metals by changing the top layer tungsten and GP tungsten with other metals (and refractory ceramic TiN) without changing any other dimensions. The results are summarized in Table.4 and the response of these absorbers for visible regime is plotted in Fig.16.

Table 4: Absorbance by different metals and refractory ceramic

| Metal | Curve fitting (max. Coefficients) | Absorbance |
|---|---|---|
| Silver (Ag) | 6 | 36.3412 |
| Gold (Au) | 9 | 65.255 |
| Copper (Cu) | 8 | 68.4969 |
| Iron (Fe) | 6 | 96.6325 |
| Titanium Nitride (TiN) | 6 | 92.1098 |
| Tungsten (W) | 15 | 99.3141 |

The major principle of absorbance in other metals and TiN is impedance matching and surface plasmonic resonance [19, 20]. This however, is not true for tungsten as this high absorbance is achieved solely on impedance matching. The results show that W based configuration is better optimized as an absorber, though Fe is close. Noble metals such as Ag and Au are not giving as better response as W for the cross based structure of a PLA.

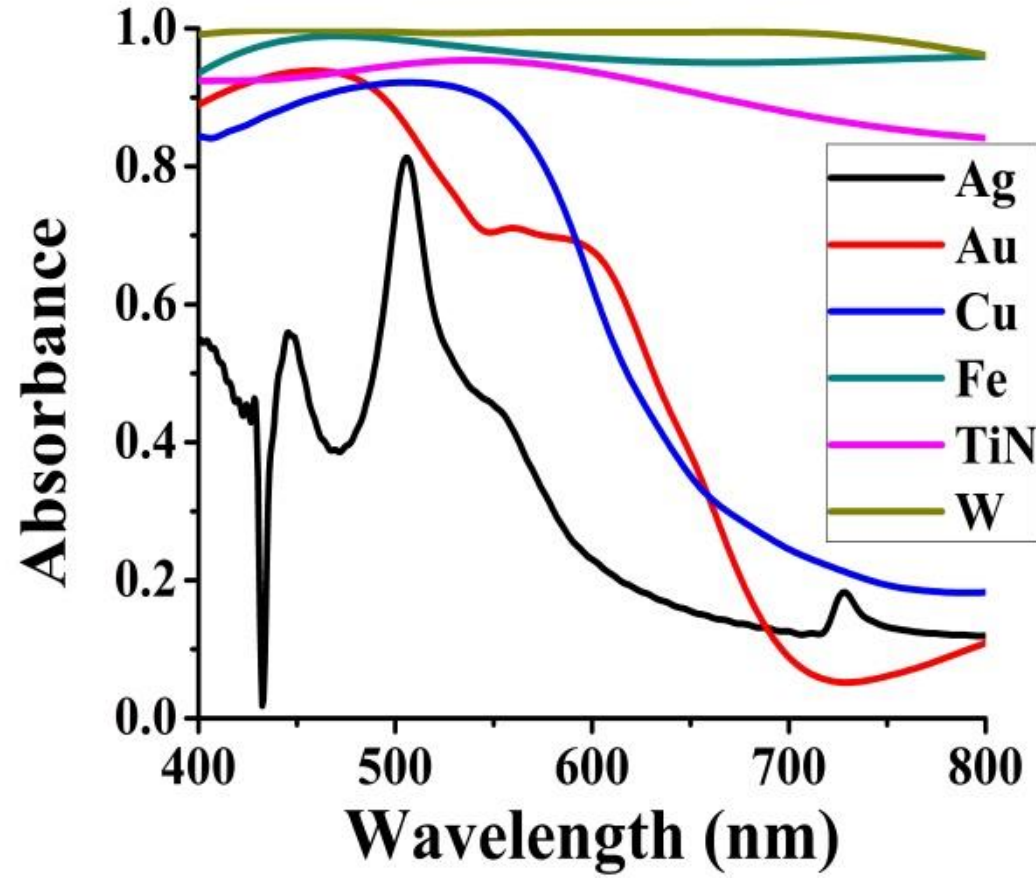

Figure 16: Absorbance achieved by various metals and TiN.

## CONCLUSION

It can be concluded that cross design is highly optimized for Tungsten. This design is consistent in z-direction and only varies in x and y directions. Therefore, this device can be called as two-dimensional (2D). As cross is symmetrical it is polarization insensitive. The high absorbance achieved by this design is because it is impedance matched with free space. This absorber gives almost constant response for sources which are incident at an angle theta (θ) < 70°. Even at 80°it does not lose all its absorbance capability but diminishes to about 50% (Fig.14 and Fig.15). Tungsten based absorber also outperforms absorbers based on other metals.

## REFERENCES

[1] N. I. Landy, S. Sajuyigbe, J. Mock, D. Smith, and W. Padilla, "Perfect metamaterial absorber," *Physical review letters,* vol. 100, p. 207402, 2008.

[2] H. Tao, N. I. Landy, C. M. Bingham, X. Zhang, R. D. Averitt, and W. J. Padilla, "A metamaterial absorber for the terahertz regime: Design, fabrication and characterization," *Optics express,* vol. 16, pp. 7181-7188, 2008.

[3] K. Bhattarai, S. Silva, K. Song, A. Urbas, S. J. Lee, Z. Ku, and J. Zhou, "Metamaterial Perfect Absorber Analyzed by a Meta-cavity Model Consisting of Multilayer Metasurfaces," *arXiv preprint arXiv:1705.02681,* 2017.

[4] S. Cao, W. Yu, T. Wang, Z. Xu, C. Wang, Y. Fu, and Y. Liu, "Two-dimensional subwavelength meta-

nanopillar array for efficient visible light absorption," *Applied Physics Letters,* vol. 102, p. 161109, 2013.
[5] C.-W. Cheng, M. N. Abbas, C.-W. Chiu, K.-T. Lai, M.-H. Shih, and Y.-C. Chang, "Wide-angle polarization independent infrared broadband absorbers based on metallic multi-sized disk arrays," *Optics express,* vol. 20, pp. 10376-10381, 2012.
[6] M. Iwanaga, "Perfect Light Absorbers Made of Tungsten-Ceramic Membranes," *Applied Sciences,* vol. 7, p. 458, 2017.
[7] A. K. Azad, W. J. Kort-Kamp, M. Sykora, N. R. Weisse-Bernstein, T. S. Luk, A. J. Taylor, D. A. Dalvit, and H.-T. Chen, "Metasurface broadband solar absorber," *Scientific reports,* vol. 6, 2016.
[8] Q. Han, Y. Fu, L. Jin, J. Zhao, Z. Xu, F. Fang, J. Gao, and W. Yu, "Germanium nanopyramid arrays showing near-100% absorption in the visible regime," *Nano Research,* vol. 8, pp. 2216-2222, 2015.
[9] C. Ungaro, S. K. Gray, and M. C. Gupta, "Black tungsten for solar power generation," *Applied Physics Letters,* vol. 103, p. 071105, 2013.
[10] Y. Lin, Y. Cui, F. Ding, K. H. Fung, T. Ji, D. Li, and Y. Hao, "Tungsten based anisotropic metamaterial as an ultra-broadband absorber," *Optical Materials Express,* vol. 7, pp. 606-617, 2017.
[11] W. Li, U. Guler, N. Kinsey, G. V. Naik, A. Boltasseva, J. Guan, V. M. Shalaev, and A. V. Kildishev, "Refractory plasmonics with titanium nitride: broadband metamaterial absorber," *Advanced Materials,* vol. 26, pp. 7959-7965, 2014.
[12] T. M. Mattox, A. Bergerud, A. Agrawal, and D. J. Milliron, "Influence of shape on the surface plasmon resonance of tungsten bronze nanocrystals," *Chemistry of Materials,* vol. 26, pp. 1779-1784, 2014.
[13] A. A. Shah and M. C. Gupta, "Spectral selective surfaces for concentrated solar power receivers by laser sintering of tungsten micro and nano particles," *Solar Energy Materials and Solar Cells,* vol. 117, pp. 489-493, 2013.
[14] E. D. Palik, *Handbook of optical constants of solids* vol. 3: Academic press, 1998.
[15] D. Smith, S. Schultz, P. Markoš, and C. Soukoulis, "Determination of effective permittivity and permeability of metamaterials from reflection and transmission coefficients," *Physical Review B,* vol. 65, p. 195104, 2002.
[16] D. Smith, D. Vier, T. Koschny, and C. Soukoulis, "Electromagnetic parameter retrieval from inhomogeneous metamaterials," *Physical review E,* vol. 71, p. 036617, 2005.
[17] X. Chen, T. M. Grzegorczyk, B.-I. Wu, J. Pacheco Jr, and J. A. Kong, "Robust method to retrieve the constitutive effective parameters of metamaterials," *Physical review E,* vol. 70, p. 016608, 2004.
[18] A. Sellier, T. V. Teperik, and A. de Lustrac, "Resonant circuit model for efficient metamaterial absorber," *Optics express,* vol. 21, pp. A997-A1006, 2013.
[19] W. L. Barnes, A. Dereux, and T. W. Ebbesen, "Surface plasmon subwavelength optics," *nature,* vol. 424, p. 824, 2003.
[20] W. L. Barnes, "Surface plasmon–polariton length scales: a route to sub-wavelength optics," *Journal of optics A: pure and applied optics,* vol. 8, p. S87, 2006.